\documentclass{elsart}

\usepackage[tbtags]{amsmath}
\usepackage{amscd}
\usepackage{amsfonts}
\usepackage{epsfig}
\usepackage{comment}

\renewcommand{\mod}{\;\mbox{mod}\;}
\renewcommand{\vec}[1]{\mbox{\boldmath $#1$}}
\newcommand{\mat}[1]{{\mbox{\bfseries \rmfamily#1}}}
\newcommand{\Idmat}{\vec{1}}
\newcommand{\diag}{{\rm diag}}

\newcommand{\mut}{R}

\allowdisplaybreaks

\begin{document}
\begin{frontmatter}

\title{Dynamic fitness landscapes: Expansions for small mutation rates}

\author{Claus O.\ Wilke}
\address{
Digital Life Laboratory\\
Mail Code 136-93, Caltech\\
Pasadena, CA 91125\\
wilke@caltech.edu
}

\author{Christopher Ronnewinkel}
\address{
Institut f\"ur Neuro- und Bioinformatik\\
Medizinische Universit\"at L\"ubeck\\
Seelandstra\ss e 1a\\
D-23569 L\"ubeck, Germany\\
ronne@inb.mu-luebeck.de
}


\begin{abstract}
  We study the evolution of asexual microorganisms with small mutation rate in
  fluctuating environments, and develop techniques that allow us to expand the
  formal solution of the evolution equations to first order in the mutation
  rate. Our method can be applied to both discrete time and continuous time
  systems. While the behavior of continuous time systems is dominated by the
  average fitness landscape for small mutation rates, in discrete time systems
  it is instead the geometric mean fitness that determines the system's
  properties. In both cases, we find that in situations in which the
  arithmetic (resp.\ geometric) mean of the fitness landscape is degenerate,
  regions in which the fitness fluctuates around the mean value present a
  selective advantage over regions in which the fitness stays at the mean.
  This effect is caused by the vanishing genetic diffusion at low mutation
  rates. In the absence of strong diffusion, a population can stay close to a
  fluctuating peak when the peak's height is below average, and take
  advantage of the peak when its height is above average.
\end{abstract}

\end{frontmatter}

\section{Introduction}
A major part of all living creatures on Earth consists of prokaryotes and
phages. These organisms replicate mainly without sexual
recombination~\cite{LevinBergstrom2000}, and typically produce offspring on a
time-scale of hours. Because of their short gestation times, microbes
experience ubiquitous environmental changes such as seasons on an evolutionary
time scale.  Most of the DNA based microbes have developed error correction
mechanisms, such limiting the amount of deleterious mutations they experience.
In a changing environment, however, small mutation rates might severely
curtail a species' ability to react to new situations. The observed genomic
mutation rates of asexual organisms such as bacteria and DNA viruses lie
typically around $2-4\times 10^{-3}$~\cite{Drakeetal98}, implying that a few
out of every thousand offspring get mutated at all.  It has been
proposed~\cite{NilssonSnoad2000b} that even lower genomic mutation rates are
not observed simply because they would stifle a species' adaptability in a
changing environment. While this is certainly a reasonable assumption, we do
not currently have a deep understanding of what types of fitness landscapes
require what mutation rates, and whether a small mutation rate is always
disadvantageous in a changing environment. In this paper, we address the
effects of a changing environment on a population evolving in a small mutation
rate. Our main objective is to develop an expansion to first order in the
mutation rate which enables us to find approximate
solutions for infinite asexual populations evolving in arbitrary dynamic
landscapes.

Due to the nature of the expansions that we use, we are led to a comparison
between discrete time and continuous time systems. Our main result from the
comparison is that in dynamic fitness landscapes, continuous and discrete time
systems have qualitative differences in the low mutation rate regime.  This
difference can manifest itself, for example, in populations that replicate
either continuously or synchronized in discrete generations. Given that all
other factors are equal, the continuously replicating strains will have a
selective advantage.  As a generic result for both continuous
and discrete time, we find that a low mutation rate can enable a
population to draw a selective advantage from fluctuations in the landscape.

Our analysis is based on the quasispecies
model~\cite{Eigen71,Eigenetal88,Eigenetal89}. The quasispecies literature was
for a long time focused on static fitness landscapes, but recently more
emphasis has been put on the aspect of changing
environments~\cite{NilssonSnoad2000b,Jones79a,Jones79b,Wilke99,WilkeRonnewinkelMartinetz99,NilssonSnoad2000,WilkeRonnewinkelMartinetz99a,RonnewinkelWilkeMartinetz2000,NilssonSnoad2000a}.
Here, we mainly use methods developed in
Ref.~\cite{WilkeRonnewinkelMartinetz99a}.  The paper is structured as follows.
In Sec.~\ref{sec:analysis}, we demonstrate how systems with discrete as well
as continuous time can be treated to first order in the mutation rate. In
Sec.~\ref{sec:discussion}, we discuss the expansions we have found in
Sec.~\ref{sec:analysis}. We treat the case of a vanishing mutation rate in
Sec.~\ref{sec:zero-mutrate}, and that of a very small but positive mutation
rate in Sec.~\ref{sec:nonzero-mutrate}. In Sec.~\ref{sec:localization}, we
study the localization of a population around a oscillating peak, and in
Sec.~\ref{sec:discrete-overlap}, we discuss the problems we encounter when
approximating a continuous time system with a discrete time system. We close
our paper with concluding remarks in Sec.~\ref{sec:conclusions}.

\section{Analysis}\label{sec:analysis}

\subsection{The model}

Consider a system of evolving bitstrings. The different bitstrings $i$
replicate with rates $A_i$, and they mutate into each other with
probabilities $Q_{ij}$.  Throughout this paper, we assume that the probability
of an incorrectly copied bit is uniform over all strings, and denote this
probability by $\mut$. The mutation matrix $\mat Q=(Q_{ij})$ is then given
by
\begin{equation}
  Q_{ij} = (1-\mut)^l \left(\frac{\mut}{1-\mut}\right)^{d(i,j)}\,,
\end{equation}
where $d(i,j)$ is the Hamming distance between two sequences $i$ and $j$. The
matrix $\mat Q$ is a $2^l\times 2^l$ matrix, and it is in general difficult to
handle numerically. Therefore, in the following we impose the additional
assumption that all sequences with equal Hamming distance from a given
reference sequence have the same fitness. This is the so-called {error class}
assumption~\cite{SwetinaSchuster82}. The matrix $\mat Q$ is then an
$(l+1)\times(l+1)$ matrix,
\begin{equation}\label{eq:error-class-Q}
  Q_{ij} = \sum_{k=\max\{i+j-l,0\}}^{\min\{i,j\}} \binom{j}{k}\binom{l-j}{i-k}
   (1-\mut)^l\left(\frac{\mut}{1-\mut}\right)^{i+j-2k}\,.
\end{equation}
The generality of our results is not affected by this choice, because the
calculations we present in the following can be performed with either of the
two matrices $\mat Q$, and they lead to very similar expressions.

Let us write down the quasispecies equations for sequences evolving in
continuous or discrete time and in a static fitness landscape. We
introduce the replication matrix $\mat A=\diag(A_0, A_1, \dots)$. The
continuous differential equation of the (unnormalized) concentration variables
$\vec y=(y_0, y_1, \dots)$ then reads
\begin{equation}\label{eq:static-differential-eq}
  \dot{\vec y}(t) = \mat Q\mat A \vec y(t)\,.
\end{equation}
The discrete difference equation, on the other hand, can be written as
\begin{equation}\label{eq:static-difference-eq}
  \vec y(t+\Delta t) = [\Delta t\mat Q\mat A + \lambda]\vec y(t)\,,
\end{equation}
where $\Delta t$ is the duration of one generation, and $\lambda$ gives the
proportion of parents that survive one generation and enter the next one
together with their offspring. Both Eq.~\eqref{eq:static-differential-eq} and Eq.~\eqref{eq:static-difference-eq} converge for
$t\rightarrow\infty$ towards a sequence distribution given by the Perron
eigenvector of the matrix $\mat Q\mat A$. Hence, for a static landscape the
discrete time and the continuous time quasispecies equations are equivalent,
as far as the asymptotic state is concerned. The distinction between discrete
and continuous time, however, is important when the
fitness landscape changes over time. Consider the situation of
a dynamic fitness landscape, represented by a time dependent matrix $\mat
A(t)$. Equation~\eqref{eq:static-differential-eq} becomes
\begin{equation}\label{eq:dynamic-differential-eq}
  \dot{\vec y}(t) = \mat Q\mat A(t) \vec y(t)\,.
\end{equation}
The time-dependent difference equation, on the other hand, reads
\begin{equation}\label{eq:dynamic-difference-eq}
  \vec y(t+\Delta t) = [\Delta t\mat Q\mat A(t)+\lambda]\vec y(t)\,,
\end{equation}
The dynamic attractors of both Eqs.~\eqref{eq:dynamic-differential-eq}
and~\eqref{eq:dynamic-difference-eq} are not immediately obvious, and
therefore we cannot know to what extent the two systems differ unless we
perform a more elaborate analysis.  Moreover, in a static landscape, a nonzero
$\lambda$ does not affect the asymptotic state of the system, which is why it
normally is set to zero in
Eq.~\eqref{eq:static-difference-eq}~\cite{Demetriusetal85,BaakeGabriel99}. The
situation is different in a dynamic landscape, and we have to allow for a
non-zero $\lambda$ in general.

\subsection{Discrete time}

Let us begin our analysis with the discrete system. We set $\lambda=0$,
because that leads to the simplest
equation describing a discrete time evolutionary system in a dynamic fitness
landscape. The more complicated cases with $\lambda>0$ can be constructed from
the equation for $\lambda=0$, as we will see later on. We
address the equation
\begin{equation}
  \vec y(t+\Delta t) = \Delta t\mat Q\mat A(t)\vec y(t)\,.
\end{equation}
The solution to this equation is formally given by the time-ordered matrix
product~\cite{WilkeRonnewinkelMartinetz99a} [using $n=t/\Delta t$ and $\mat
A'(\nu)=\Delta t\,\mat A(\nu\Delta t)$]
\begin{align}\label{eq:discrete-formal-sol}
  \vec y(n) &=\mathcal{T}\left\{\prod_{\nu=0}^{n-1}
     \mat Q\mat A'(\nu)\right\}
  \vec y(0)  \notag\\
  &=: \mat Y_{\rm disc}(n) \vec y(0)\,.
\end{align}
In the second line, we have introduced the notation $\mat Y_{\rm disc}(n)$ for
this matrix product. We will occasionally refer to $\mat Y_{\rm disc}(n)$ as
a \emph{propagator}, since  $\mat Y_{\rm disc}(n)$ fully determines the
state of the system at time $t=n\,\Delta t$, given an initial state at time $t=0$.

 $\mat Y_{\rm disc}(n)$ can be evaluated to first order in
$\mut$. The only dependency of $\mat Y_{\rm disc}(n)$ on $\mut$ is the one in
$\mat Q$. When we expand $\mat Q$ [Eq.~\eqref{eq:error-class-Q}] in $\mut$, we
find
\begin{equation}\label{eq:error-class-Q-expans}
  Q_{ij} = \sum_{k=\max\{i+j-l,0\}}^{\min\{i,j\}} \binom{j}{k}\binom{l-j}{i-k}
   [\delta_{\beta,0} +\mut(\delta_{\beta,1}-\alpha\delta_{\beta,0})+\dots]\,,
\end{equation}
with $\alpha=l-\beta$ and $\beta=i+j-2k$. As usual, $\delta_{i,j}$ denotes the
Kronecker symbol. The sum collapses into a single term, and we find to first
order in $\mut$
\begin{equation}\label{eq:Q-expans}
  Q_{ij} = (1-l\mut)\delta_{i,j} + (l-j)\mut\delta_{i,j+1} + j\mut\delta_{i,j-1}\,.
\end{equation}
After some algebra, we obtain from that for the matrix $\mat Y_{\rm disc}(n)$
\begin{align}\label{eq:discrete-Y-expans}
  \Big(\mat Y_{\rm disc}(n)\Big)_{ij} &= \left[(1-l\mut n)
    \prod_{\nu=0}^{n-1} A'_j(\nu)\right]\delta_{i,j} \notag\\
  &\qquad + \left[(l-j)\mut \sum_{\mu=0}^{t-1} \prod_{\nu_1=0}^{\mu} A'_j(\nu_1)
  \prod_{\nu_2=\mu+1}^{t-1} A'_{j+1}(\nu_2)\right] \delta_{i,j+1} \notag\\
  &\qquad + \left[ j\mut \sum_{\mu=0}^{t-1} \prod_{\nu_1=0}^{\mu} A'_j(\nu_1)
  \prod_{\nu_2=\mu+1}^{t-1} A'_{j-1}(\nu_2)\right] \delta_{i,j-1}\,,
\end{align}
This expression fully describes to first order in $\mut$ the state of the system
after $n$ time steps.

\subsection{Continuous time}

Let us now turn to the continuous system. We can use the expansion of $\mat
Y_{\rm disc}(n)$ to find an expansion for the propagator of the static
continuous case. If the
fitness landscape is static, the solution to
Eq.~\eqref{eq:static-differential-eq} is given by
\begin{equation}
  \vec y(t) = \exp(\mat Q\mat A t) \vec y(0)\,.
\end{equation}
It is useful to recall that the exponential operator of a matrix is defined as
\begin{equation}\label{eq:exp-expans}
  \exp(\mat Q\mat A t) = \Idmat + \mat Q\mat A t + \frac{1}{2!}
       (\mat Q\mat A t)^2 + \dots\,.
\end{equation}
We can expand the single terms in that sum separately.
Equation~\eqref{eq:discrete-Y-expans} allows us to write $(\mat Q\mat A)^k$ as
\begin{align}
  \Big((\mat Q\mat A)^k\Big)_{ij} &= (1-kl\mut)A_j^k\delta_{i,j} +
  \left[(l-j)\mut\sum_{\mu=1}^k A_j^\mu A_{j+1}^{k-\mu}\right]
  \delta_{i,j+1} \notag\\
  &\qquad +\left[j\mut\sum_{\mu=1}^k  A_j^\mu A_{j-1}^{k-\mu}\right]
    \delta_{i,j-1} + \mathcal{O}(\mut^2)\,.
\end{align}
Both the sums in the expression for $(\mat Q\mat A)^k$ and the remaining sum
in Eq.~\eqref{eq:exp-expans} can then be taken analytically. We find
\begin{align}\label{eq:static-cont-expans}
  \Big(\exp(\mat Q\mat A t)\Big)_{ij} &= (1-l\mut A_j t)e^{A_j t}\delta_{i,j} +
  (l-j)\mut K[A_j t, A_{j+1}t] \delta_{i,j+1} \notag\\
  &\qquad +j\mut K[A_j t,A_{j-1}t]
    \delta_{i,j-1}\,,
\end{align}
where the function $K[a,b]$ is of the form
\begin{equation}
  K[a,b] = \frac{a}{a-b}\left(e^{a} - e^{b}\right).
\end{equation}
With Eq.~\eqref{eq:static-cont-expans}, we have an expansion of the
propagator of the continuous system in a static landscape to first order in
$\mut$. Similarly, we can treat piecewise constant landscapes. Under a
piecewise constant landscape we understand a landscape for which we can
define intervals $I_1=[0,t_1)$, $I_2=[t_1,t_2)$, $I_3=[t_2,t_3)$, $\dots$,
such that the landscape does not change within any of these intervals.
Any dynamic fitness landscape can be approximated in that way. The
solution to the differential equation for that type of landscapes is given
by
\begin{align}\label{eq:sol-piecewise-const}
  \vec y(t) &= \exp[\mat Q\mat A(t_n)(t-t_n)]\exp[\mat Q\mat A(t_{n-1})(t_n-t_{n-1})] \cdots\notag\\
  &\qquad\cdots \exp[\mat Q\mat A(0)t_1] \vec y(0)\,.
\end{align}
With the two simplifying assumptions that all intervals have the same
length $\tau$ and that we are observing the system only at the end of an
interval, Eq.~\eqref{eq:sol-piecewise-const} becomes (for $n=t/\tau$)
\begin{align}
  \vec y(t) &=  \mathcal{T}\left\{\prod_{\nu=0}^{n-1} \exp[\mat Q\mat A(\nu\tau)\tau]\right\} \vec y(0) \notag\\
  &=: \mat Y_{\rm cont}(t) \vec y(0)\,.
\end{align}
The similarity to Eq.~\eqref{eq:discrete-formal-sol} is evident. Hence,
in analogy to the calculation that leads from Eq.~\eqref{eq:Q-expans}
to Eq.~\eqref{eq:discrete-Y-expans}, we find in the piecewise
constant, continuous case
\begin{multline}\label{eq:continuous-Y-expans}
  \Big(\mat Y_{\rm cont}(t)\Big)_{ij} = \left[\Big(1-l\mut
         \sum_{\nu=0}^{n-1}A_j^{c}(\nu)\Big)
       \exp\Big(\sum_{\nu=0}^{n-1}A_j^{c}(\nu)\Big)\right]\delta_{i,j}\\
  \qquad + \left[(l-j)\mut \sum_{\mu=0}^{n-1}K^{(\mu)}_{j,j+1}
     \prod_{\nu_1=0}^{\mu-1} \exp[A_j^{c}(\nu_1)]
  \prod_{\nu_2=\mu+1}^{n-1}\exp[A_{j+1}^{c}(\nu_2)]\right] \delta_{i,j+1}\\
  \qquad + \left[ j\mut \sum_{\mu=0}^{n-1}K^{(\mu)}_{j,j-1}
     \prod_{\nu_1=0}^{\mu-1} \exp[A_j^{c}(\nu_1)]
  \prod_{\nu_2=\mu+1}^{n-1}\exp[A_{j-1}^{c}(\nu_2)]\right] \delta_{i,j-1}\,.
\end{multline}
We have used the abbreviations
\begin{equation}
\mat A^{c}(\nu)=\tau\mat A(\nu\tau)\quad
\mbox{and}\quad K^{(\nu)}_{i,j} = K[A_i^{c}(\nu), A_{j}^{c}(\nu)].
\end{equation}
Equation \eqref{eq:continuous-Y-expans} fully determines to first order in
$\mut$ the state of the continuous system after $t$ units of time have passed.

\section{Discussion}\label{sec:discussion}

With Eqs.~\eqref{eq:discrete-Y-expans} and \eqref{eq:continuous-Y-expans},
we have expansions for the propagators of discrete-time and continuous time
evolving systems in a dynamic fitness landscape. In this section, we will
examine these expansions and discuss their properties.

\subsection{A vanishing mutation rate}\label{sec:zero-mutrate}

For $\mut=0$, both $\mat Y_{\rm disc}(t)$ and $\mat Y_{\rm cont}(t)$ turn into
diagonal matrices. We find (choosing $\tau=\Delta t$ and $n=t/\tau$)
\begin{align}\label{eq:zero-order-discrete}
  Y_{\rm disc}(t) & = \exp\big[ \sum_{\nu=0}^{n-1}
\log\big(\mat A'(\nu)\big)\big] = \left(\!\sqrt[\scriptstyle n]{
\mat A'(n-1)\cdots\mat A'(0)}\,\right)^{n}\,,\\\label{eq:zero-order-continuous}
  Y_{\rm cont}(t) &= \exp\big[\sum_{\nu=0}^{n-1}\mat A^{c}(\nu)\big]\,.
\end{align}
This result shows that in a dynamic fitness landscape the discrete and the
continuous model have not only quantitative, but also important qualitative
differences.  While in the continuous case the state of the system at time $t$
is determined by the exponential of the arithmetic mean of the fitness
landscape until time $t$, in the discrete case it is determined by the
exponential of the geometric mean of the fitness landscape, which can be
written as arithmetic mean of the logarithm of the fitness landscape. The
latter corresponds to results from population
genetics~\cite{YoshimuraJansen96}.  Since arithmetic and geometric mean are in
general different, the same fitness landscape can have very different effects
in a continuous or discrete system for $\mut=0$.  Consider a landscape, for
example, with an oscillating sharp peak,
\begin{subequations}\label{eq:fitness-landscape}
\begin{align}
A_{0}(t) &= \begin{cases}\sigma(1-a) \qquad \mbox{for $\phantom{T/}0\le t < T/2$}\\
\sigma(1+a) \qquad \mbox{for $T/2\le t < T$}\end{cases}\\
A_{i}(t) &= 1 \qquad \mbox{for $0<i\le l$}\,,
\end{align}
\end{subequations}
with $0\le a<1$ and $\sigma>0$.

\begin{figure}
\centerline{
 \epsfig{file={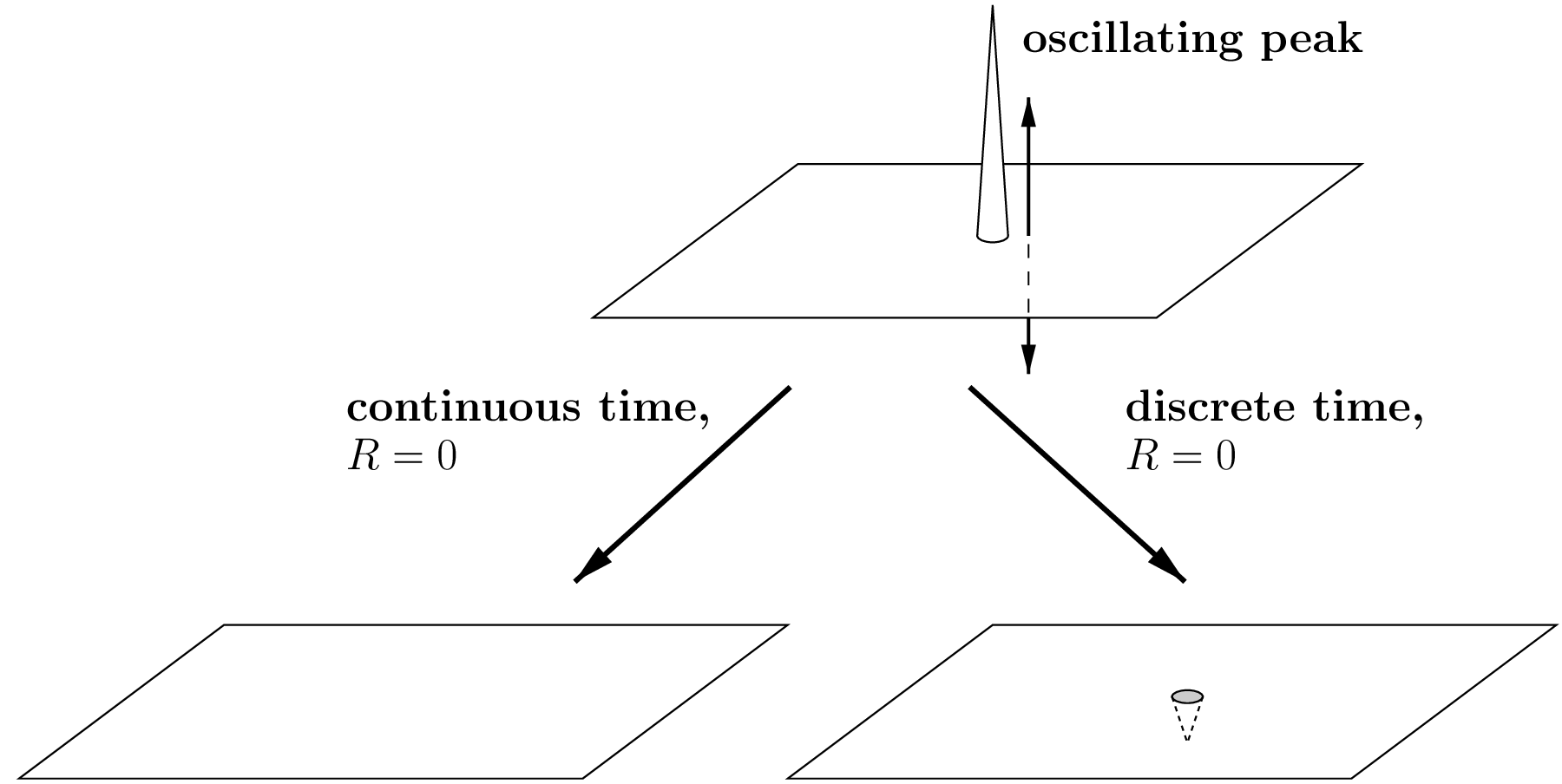}, width=11cm}
}
\caption{A landscape with an oscillating peak whose average height
  coincides with the fitness of all other sequences. In continuous time,
  the landscape becomes completely flat for $\mut = 0$. In discrete time,
  however, the population feels the geometric mean of the fitness
  landscape for $\mut = 0$, which has a hole at the position of the
  peak.}\label{fig:oscpeak}
\end{figure}

In the continuous system without mutations, the master sequence grows with the
rate $\overline{A}_{0}=\sigma$ if time is measured in integer multiples of
$T$. Hence, if $\sigma>1$, the peak sequence will always supersede all other
sequences for $t\to\infty$. Contrasting to that, the geometric mean is
$\widetilde{A}_{0}=\sigma\sqrt{1-a^{2}}$.  Even for $\sigma>1$ it is possible
to have $\widetilde{A}_{0}<1$ if $a$ is large enough, in which case in the
discrete system the master sequence grows slower than all others.
Consequently, it will be expelled from the population for $t\to\infty$. The
special case of $\sigma=1$ is depicted in Fig.~\ref{fig:oscpeak}. There,
the fitness landscape becomes flat in continuous
time, but acquires a hole in discrete time.

\subsection{Small non-zero mutation rates}\label{sec:nonzero-mutrate}

Let us now turn to the case of a small but non-zero $\mut$. From the above, we
can expect that there is a qualitative difference between discrete and
continuous time even for finite $R$.  In order to see this difference, we take
the oscillating sharp peak landscape as a generic example. A two concentration
approximation has proven useful to describe situations with $\sigma\gg
1$~\cite{NilssonSnoad2000a} but is not applicable here, since we are
particularly interested in the case $\sigma=1$, for which the average
landscape is flat in continuous time and acquires a hole in discrete time.

The analysis of the landscape Eq.~\eqref{eq:fitness-landscape} is facilitated
by its periodicity in time (with period length $T$).  For periodic landscapes,
it has been shown in Ref.~\cite{WilkeRonnewinkelMartinetz99a} that a periodic
attractor with period length $T$ exists. Its state at phase $\phi=0$ (the
phase is defined as $\phi:=t\mod T$) is given by the principal eigenvector of
the \emph{monodromy} matrix
\begin{equation}\label{eq:monodromy}
  \mat X(0) = \mat Y(T)\,,
\end{equation}
where $\mat Y(t)$ is the propagator of the system.
Equation~\eqref{eq:monodromy} holds regardless of continuous or discrete
time. The attractor's state at other phases $\phi$ can be calculated in a
similar fashion.

\begin{figure}
\centerline{
 \epsfig{file={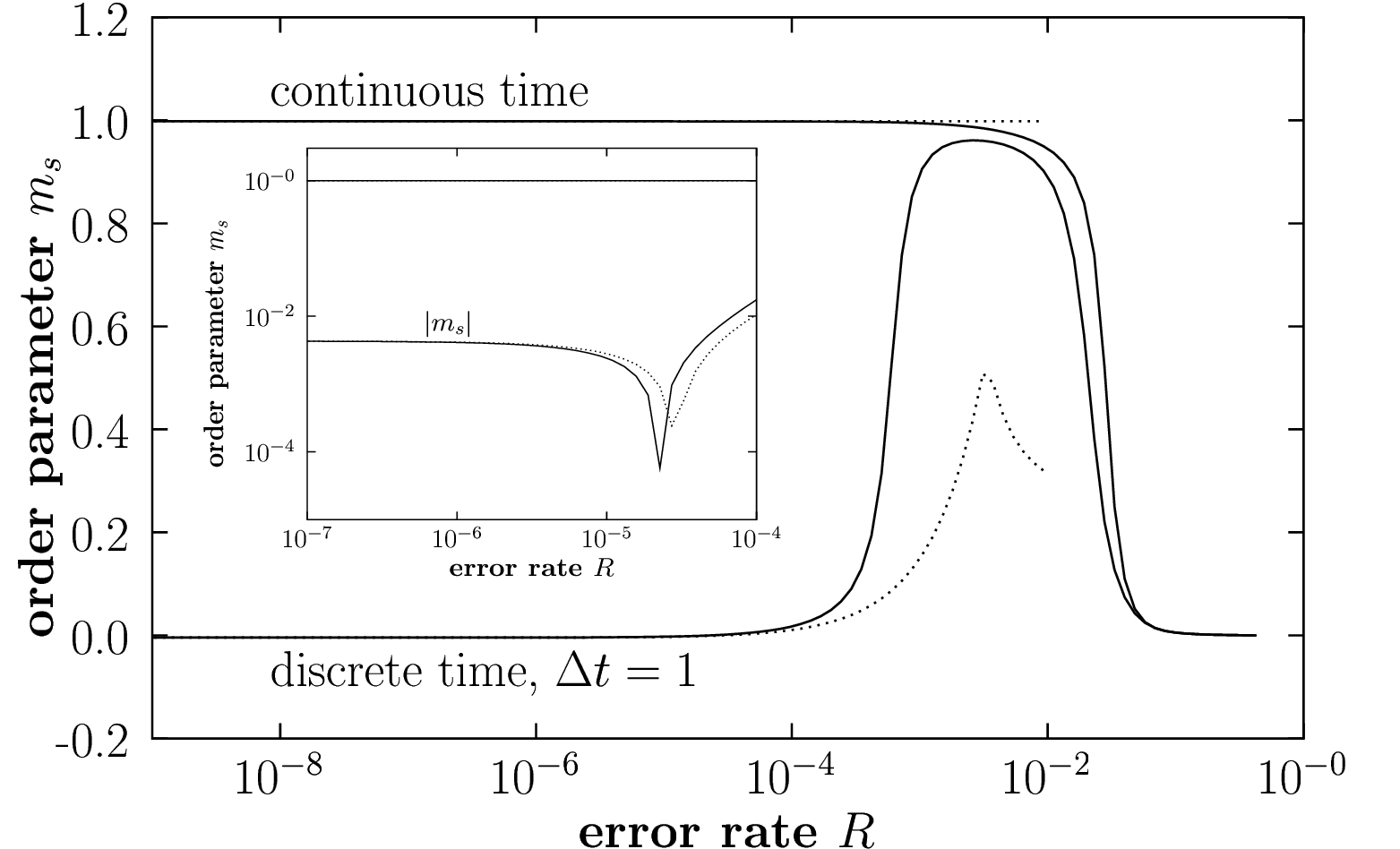}, width=11cm}
}
\caption{Order parameter $m_s$ in a dynamic fitness landscape with a single
  oscillating peak in a continuous time system and a discrete time system.
  The solid lines have been obtained from diagonalizing the full monodromy
  matrix $\mat X$, the dotted lines represent the approximation to first
  order in $\mut$. We have used the fitness landscape defined in
  Eq.~\eqref{eq:fitness-landscape}, with $a=8/10$, $T=30$, and $l=10$. The
  graph shows a snapshot of the order parameter at phase $\phi=0$ of its
  limit cycle. The inset shows the same data, but in a log-log plot. There, we
  have plotted the absolute value of $m_s$ for the discrete-time system,
  because $m_s$ assumes a value slightly below 0 in that
  case.}\label{fig:app-vs-ex}
\end{figure}

In Figure \ref{fig:app-vs-ex}, we have displayed the \emph{order parameter}
$m_s$~\cite{Leuthaeusser87,Tarazona92} in the sharp peak landscape as a
function of $R$ for the discrete time and the continuous time system. The
order parameter is given by
\begin{equation}
  m_s(t) = \frac{1}{l}\sum_{i=0}^{l} x_i(t)(l-2i)\,,
\end{equation}
where the $x_i(t)$ represent the total (normalized) concentration of all
sequences in error class $i$ at time $t$. We have calculated the order
parameter both from the full monodromy matrix and from the expansions to first
order in $\mut$. We find that the expansions give reliable results for small
mutation rates, but start deviating from the true value as $\mut$ approaches
$1/lT$. Note that both expansions must break down beyond $1/lT$, as
both the discrete and the continuous propagator assume unphysical negative
values on the diagonal when $\mut$ exceeds $1/lT$
[Eqs.~\eqref{eq:discrete-Y-expans} and \eqref{eq:continuous-Y-expans}].

From Fig.~\ref{fig:app-vs-ex}, it is evident that there exists a qualitative
difference between the discrete and the continuous time system. In the system
with continuous time, the sequences stay centered around the currently active
peak for arbitrarily small but non-zero mutation rates, whereas in the system
with discrete time, the sequence distribution becomes ever more homogeneous as
$\mut\rightarrow 0$.

The behavior of the discrete system is easily explained. In the geometric mean
of the landscape, the peak position is actually disadvantageous, and hence the
population is driven into the remaining genotype space, which it occupies
homogeneously due to the lack of selective differences. Formally, the
population feels the geometric mean only for a vanishing mutation rate.
However, by continuity, the disadvantage at the peak position will remain for
some small but non-zero $\mut$, which leads to the continuous decay of the
order parameter as $\mut \rightarrow 0$. Interestingly, the order parameter
does not decay exactly to zero, but to a value slightly below zero.  This
happens because the population becomes homogeneously distributed over the
whole sequence space except for the position of the oscillating peak. The
resulting small imbalance in the sequence distribution towards the opposite
end of the boolean hypercube then leads to a negative order parameter. The
inset in Fig.~\ref{fig:app-vs-ex} shows that our approximation predicts this
behavior accurately for small $\mut$.

Now consider the continuous system. For an infinitesimal $R>0$, the dependence
of the asymptotic state on the initial condition is lost, as we know from the
Frobenius-Perron theorem. Since for $R=0$ the evolution of the population in
time steps of size $T$ is guided by the flat average landscape, one might
suspect that for infinitesimal $R>0$ a homogeneous distribution is found as
the unique asymptotic state. This is what we observe for a population evolving
in a flat static landscape with little mutation.  However,
the situation in a dynamic landscape may be different, because the dynamics of the landscape has a significant influence on
the asymptotic sequence distribution. In fact, it is possible that a flat
average landscape leads to an ordered asymptotic state for finite mutation
rates $\mut>0$. In the next subsection, we will demonstrate this effect for the
oscillating sharp peak.

\subsection{Localization around an oscillating peak}
\label{sec:localization}

We will now have a closer look at the oscillating sharp peak landscape,
Eq.~\eqref{eq:fitness-landscape}. We are interested in the case $\sigma=1$,
which leads to a degenerate average in continuous time. First we note some
general properties of the monodromy matrix $\mat X(\phi)$  for a
periodic landscape with flat arithmetic mean.  If $\mat X(\phi)$ is given to
first order in $R$, it reads (assuming the average fitness is 1)
\begin{equation}
\mat X(\phi) = (1 - T l \mut)\mat 1 + \mut\tilde{\mat X}(\phi)\,,
\end{equation}
where $\tilde{\mat X}(\phi)$ is independent of the mutation rate $\mut$ and
contains only the off-diagonal entries from
Eq.~\eqref{eq:continuous-Y-expans}. Since $\tilde{\mat X}(\phi)$ differs
from $\mat X(\phi)$ only by a scalar factor and an additional constant on
the diagonal, the eigenvectors of the former matrix are identical to
the ones of the latter matrix, while the eigenvalues are related
through $\lambda_i = (1-Tl\mut) + \mut \tilde\lambda_i$.  As a
consequence, we find that the asymptotic species distribution is given
by the principal eigenvector of the off-diagonal matrix
$\tilde{\mat X}(\phi)$, which is independent of $\mut$.
If we took terms up to the $k$th order of $\mut$ into account in
Eq.~\eqref{eq:continuous-Y-expans}, we would find the higher order
contributions to the eigenvectors up to $(k-1)$th order of $\mut$. However,
with our first-order approximation, we are only able to
calculate the asymptotic sequence concentrations to 0th order in $\mut$.

For small mutation rates $\mut$, we can restrict our analysis to the first
three error classes. For the oscillating peak, we find with help
of Eq.~\eqref{eq:continuous-Y-expans} the following expressions
\begin{equation}
\tilde{\mat X}(\phi) \approx T\exp(T)\left(\begin{matrix}
0&\alpha_{\phi}(aT)&0\\
l\beta_{\phi}(aT)\alpha_{\phi}(aT)&0&2\\
0&l-1&0
\end{matrix}\right)\,,
\end{equation}
where
\begin{subequations}
\begin{align}
\alpha_{\phi}(\xi)&=(2/\xi)[1-e^{-\xi/2}]e^{\xi|\phi-1/2|}\,\\
\beta_{\phi}(\xi)&=e^{\xi/2}e^{-2\xi|\phi-1/2|}\,\\
\xi &= a T\,.
\end{align}
\end{subequations}
The (unnormalized) asymptotic state follows as
\begin{equation}\label{eq:approxqs}
\left(\begin{matrix}y_{0}\\y_{1}\\y_{2}\end{matrix}\right)(\phi,\xi)
 = \left(\begin{matrix}\alpha_{\phi}(\xi)\\
\sqrt{2(l-1)+\alpha_{\phi}^{2}(\xi)\beta_{\phi}(\xi)\,l}\\
l-1\end{matrix}\right)\,.
\end{equation}
Now, if the third error class concentration is negligibly small
compared to the other two concentrations, the concentrations of the
higher error classes can be neglected as well, and the asymptotic state
is approximately given by the concentrations of the first two error
classes only. From Eq.~\eqref{eq:approxqs}, we can derive the
following criterion for this approximation to be valid,
\begin{equation}\label{eq:approx-cond}
\exp(\xi/2)\gg l\xi^{2}/4.
\end{equation}
Hence, if $\xi=aT$ is large, which means that the fitness fluctuations are
large and slow, the population is exclusively distributed over the peak
and the first error class. For this case, we find the following simplified
description of the population:
\begin{subequations}\label{eq:approx}
\begin{align}
x_{0}(\phi,\xi)&=1\Big/\left[1+\sqrt{l\beta_{\phi}(\xi)}\right]\,,\\
x_{1}(\phi,\xi)&=\sqrt{l\beta_{\phi}(\xi)}\Big/\left[1+\sqrt{l\beta_{\phi}(\xi)}\right]
\,,\\
m_{s}(\phi,\xi)&=1-2x_{1}(\phi,\xi)/l\,.
\end{align}
\end{subequations}
The last equation implies that in the limit $R\to0$, the order parameter
is always larger than $1-2/l$. This means that although the peak does not have
an average selective advantage, the evolving sequences are attracted to the
peak nonetheless. As long as $l>1$, the order parameter averaged over one
oscillation cycle is \emph{positive}, which means that a population can draw a
selective advantage from being close to the peak in comparison to being far
away from it.

\begin{figure}
  \centerline{\epsfig{file={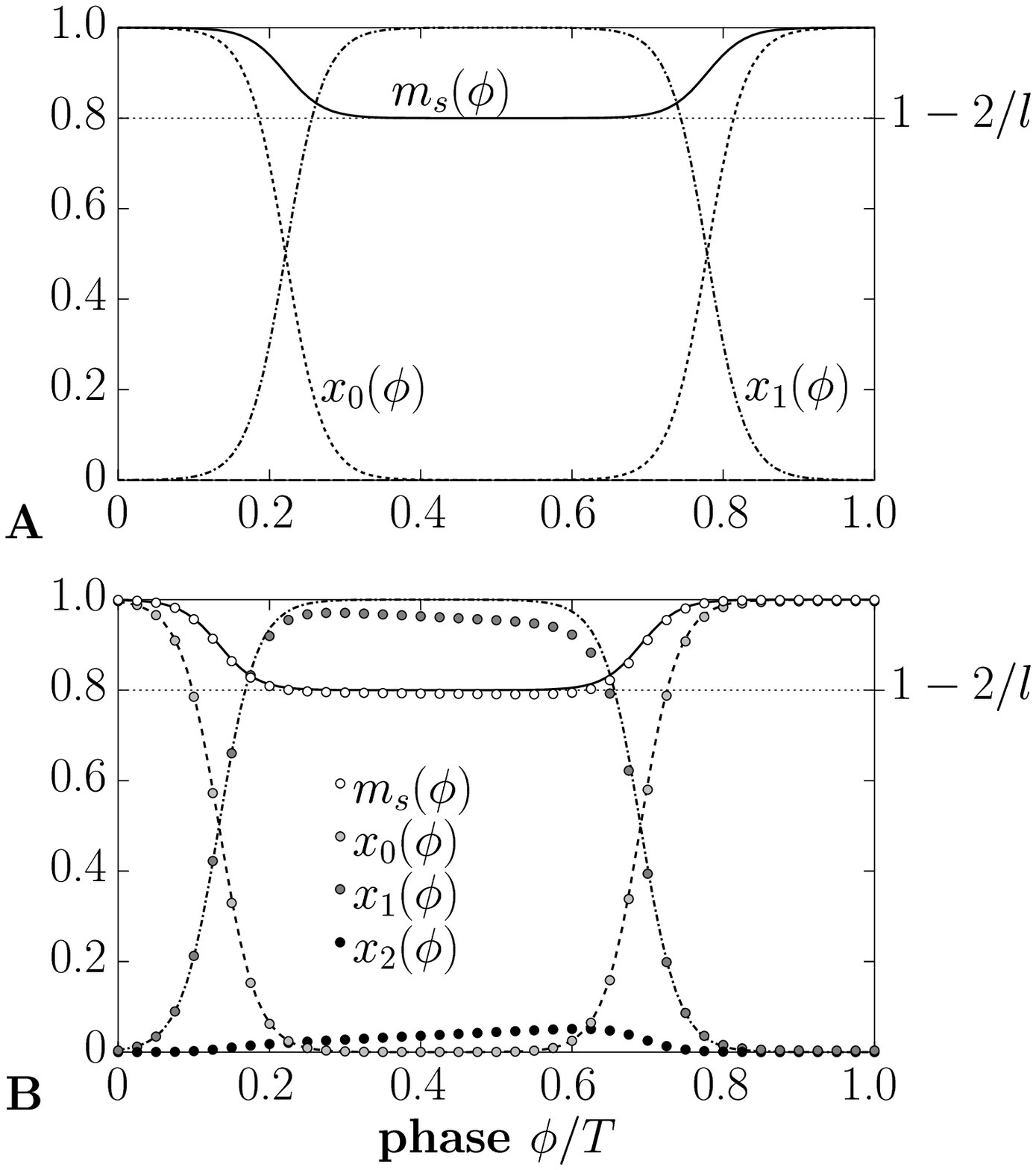}, width=9cm}}
\caption{%
  Sequence concentrations and order parameter of the steady state versus the
  relative phase $\phi/T$. The upper plot (A) shows the predictions for
  $\mut\to0$ from the three-concentration model [Eq.~\eqref{eq:approx}], with
  $T=100$, $a=0.4$ and $l=10$. The line for $x_2(\phi)$ is indistinguishable
  from the abscissa. For sufficiently small mutation rates and given
  Eq.~\eqref{eq:approx-cond}, the full numeric solution is in perfect
  agreement with the three-concentration model. For larger mutation rates, the
  main discrepancy arises as a phase shift. For $\mut=10^{-4}$ (B), the
  prediction is still in good agreement with the full numeric solution (shown
  as circles) if we phase-shift our prediction by an amount of
  $\Delta\phi/T=0.088$.}\label{fig:approx}
\end{figure}

In Figure~\ref{fig:approx}A, we display the predicted behavior of the system
in a very small mutation rate, as given by Eq.~\eqref{eq:approx}. The observed
change in the sequence concentrations is explained as follows.  During the
times at which the peak has above-average fitness, the sequences on the peak
replicate faster than all others and hence grow exponentially until the peak's
concentration saturates around one, while all off-peak sequences assume
vanishing concentrations. Similarly, during the times at which the peak has
below-average fitness, the peak's relative concentration decays, while the
population moves onto the nearest advantageous sequences, which can be found
in the first error class. The sequences in all other error classes are
adaptively neutral compared to the first error class.  Hence, the amount of
sequences that move into higher error classes is solely determined by the
mutation rate.  If the mutation rate is small enough, the diffusion among
these neutral sequences becomes negligibly small on the time scale of the peak
oscillations $T$.  Therefore, the population stays mainly within the first
error class until the peak fitness switches back to the above-average value.
Thus, we find the qualitative behavior of Eq.~\eqref{eq:approx}: In a
landscape with a large and slowly oscillating sharp peak and a small mutation
rate, the population oscillates between the peak sequence and the first error
class in the asymptotic state. In short, the population becomes localized
close to the peak.

For extremely small mutation rates, Eq.~\eqref{eq:approx} agrees perfectly
with the full numerical solution. For somewhat larger mutation rates, the main
discrepancy that arises is a phase shift between the full solution and the
approximation (Figure~\ref{fig:approx}B). The phase shift moves the
concentration curves towards earlier times, i.e., the system becomes more
responsive to the changing peak as the mutation rate increases. This is
intuitively clear. With a higher mutation rate, the first error class will
already be occupied to a larger extent when the peak switches to the
below-average value, so that the concentration of the one-mutants can grow
faster towards their equilibrium value. Similarly, when the peak switches back
to the above-average value, the peak sequences have a more favorable initial
concentration, which makes them grow faster in comparison to a lower mutation
rate.

Let us shortly extend the above argumentation to broader peaks,
like peaks with linear flanks of width $1\le w\le l$:
\begin{equation}
A_{i}(t)= \max\Big\{1,\,\frac{w-i}{w}[A_{0}(t)-1]\Big\}\quad\mbox{for all $0<i\le l$}.
\end{equation}
The sharp peak from above corresponds to a peak of width $w=1$. For
arbitrary chosen width $1\le w\le l$, the population gets transported
to the $w$th error class due to the selection pressure during the
below-average peak fitness phases. The $w$th error class is in that
case the boundary of the advantageous region. Again, if the
mutation rate is sufficiently small, diffusion can be neglected and
the population will stay in the $w$th error class until the peak
fitness switches back to the above-average value. This implies that
for peaks of width $w\ge l/2$, it is possible to have $m_{s}(\phi)\le
0$ for some intermediate oscillation phases. In particular for the
maximum width $w=l$, the order parameter $m_{s}(\phi)$ will oscillate
symmetrically around zero.

In this subsection, we have only considered continuous time systems. We have
established that in a dynamic fitness landscape with flat average, a
population can draw a selective advantage from peaks that fluctuate around the
average fitness value. The same effect will occur in a discrete time system if
we consider the geometric mean of the fitness landscape instead. In other
words, in a landscape with flat geometric mean, a population with a small
mutation rate will draw a selective advantage from a peak that fluctuates
around that geometric mean.  The origin of that effect is again the vanishing
diffusion, which causes the population to remain close to the peak when the
peak has a height below the mean.

\subsection{Discrete systems with overlapping generations}
\label{sec:discrete-overlap}

When discussing the discrete system in Sec.~\ref{sec:zero-mutrate} and~\ref{sec:nonzero-mutrate}, we have set $\lambda=0$, i.e., we have made the assumption
that every sequence can generate offspring only once, and dies before the
next generation starts to replicate. The opposite extreme is $\lambda=1$,
for which no sequence ever dies. With $\lambda=1$, a sequence can
theoretically stay infinitely long in the system (in practice, the growth
of new sequences is compensated through an out-flux of old sequences, but
that is not our concern here. The details of the out-flux do not influence
the unnormalized concentration variables $\vec y(t)$ in
Eqs.~\eqref{eq:static-differential-eq}--\eqref{eq:dynamic-difference-eq}
\cite{WilkeRonnewinkelMartinetz99a}). For $\lambda=1$,
Eq.~\eqref{eq:dynamic-difference-eq} converges to
Eq.~\eqref{eq:dynamic-differential-eq} for $\Delta t\rightarrow 0$. In
other words, for $\lambda=1$ and a small $\Delta t$,
Eq.~\eqref{eq:dynamic-difference-eq} is an approximation to
Eq.~\eqref{eq:dynamic-differential-eq}. This fact has been exploited in
Ref.~\cite{WilkeRonnewinkelMartinetz99a} in order to calculate the
continuous system numerically. However, it has not been evaluated in Ref.~\cite{WilkeRonnewinkelMartinetz99a} to what extend the
discrete approximation behaves qualitatively different from the continuous
system.

Let us briefly examine how the discrete equation with $\lambda=1$ fits into the
concepts we have developed so far. For
$\lambda=1$, the propagator $\mat Y_{\rm disc}(t)$ assumes the form
\begin{equation}\label{eq:discrete-approx-overl}
 \mat Y_{\rm disc}(t) = \mathcal{T}\left\{\prod_{\nu=0}^{n-1}
     [\Delta t \mat Q\mat A(\nu\Delta t) + 1]\right\}\,,
\end{equation}
which can be rewritten into
\begin{align}
 \mat Y_{\rm disc}(t) &= \mathcal{T}\bigg\{ \Idmat +
   \Delta t\sum_{\nu=0}^{n-1}\mat Q\mat A(\nu\Delta t)
   + \Delta t^2\sum_{\nu=0}^{n-1}\sum_{\nu'=0}^{\nu-1}
       \mat Q\mat A(\nu\Delta t) \mat Q\mat A(\nu'\Delta t)\notag\\
 &\qquad\qquad +\dots+\Delta t^n \prod_{\nu=0}^{n-1}
                   \mat Q\mat A(\nu\Delta t)\bigg\}\,.
\end{align}

With the formulae given in Section~\ref{sec:analysis}, it is possible to
expand this
expression to first order in $\mut$. Since the corresponding calculation is
tedious, and the result does not give any new insights, we omit this expansion
here. Let us just consider the zeroth order term,
\begin{align}
 \mat Y_{\rm disc}(t) &= \Idmat +
   \Delta t\sum_{\nu=0}^{n-1}\mat A(\nu\Delta t)
   + \Delta t^2\sum_{\nu=0}^{n-1}\sum_{\nu'=0}^{\nu-1}
       \mat A(\nu\Delta t) \mat A(\nu'\Delta t)\notag\\
 &\qquad\qquad +\dots+\Delta t^n \prod_{\nu=0}^{n-1}
                   \mat A(\nu\Delta t) + \mathcal{O}(\mut)\,.
\end{align}
Compare this expression to Eqs.~\eqref{eq:zero-order-discrete}
and~\eqref{eq:zero-order-continuous}. For $\lambda=1$, we neither have the
exponential of the averaged landscape, nor do we have an expression that
depends solely on the geometric mean of the landscape. We obtain a mixture
between the two cases, and the size of $\Delta t$ determines which case we are
closer to. Consequently, we obtain qualitatively wrong results from the
discrete approximation if the arithmetic and geometric mean of the landscape
differ significantly. Nevertheless, the discrepancies between the results can
be restricted to arbitrary small values of the mutation rate if we choose
$\Delta t$ small enough.

\begin{figure}
\centerline{
 \epsfig{file={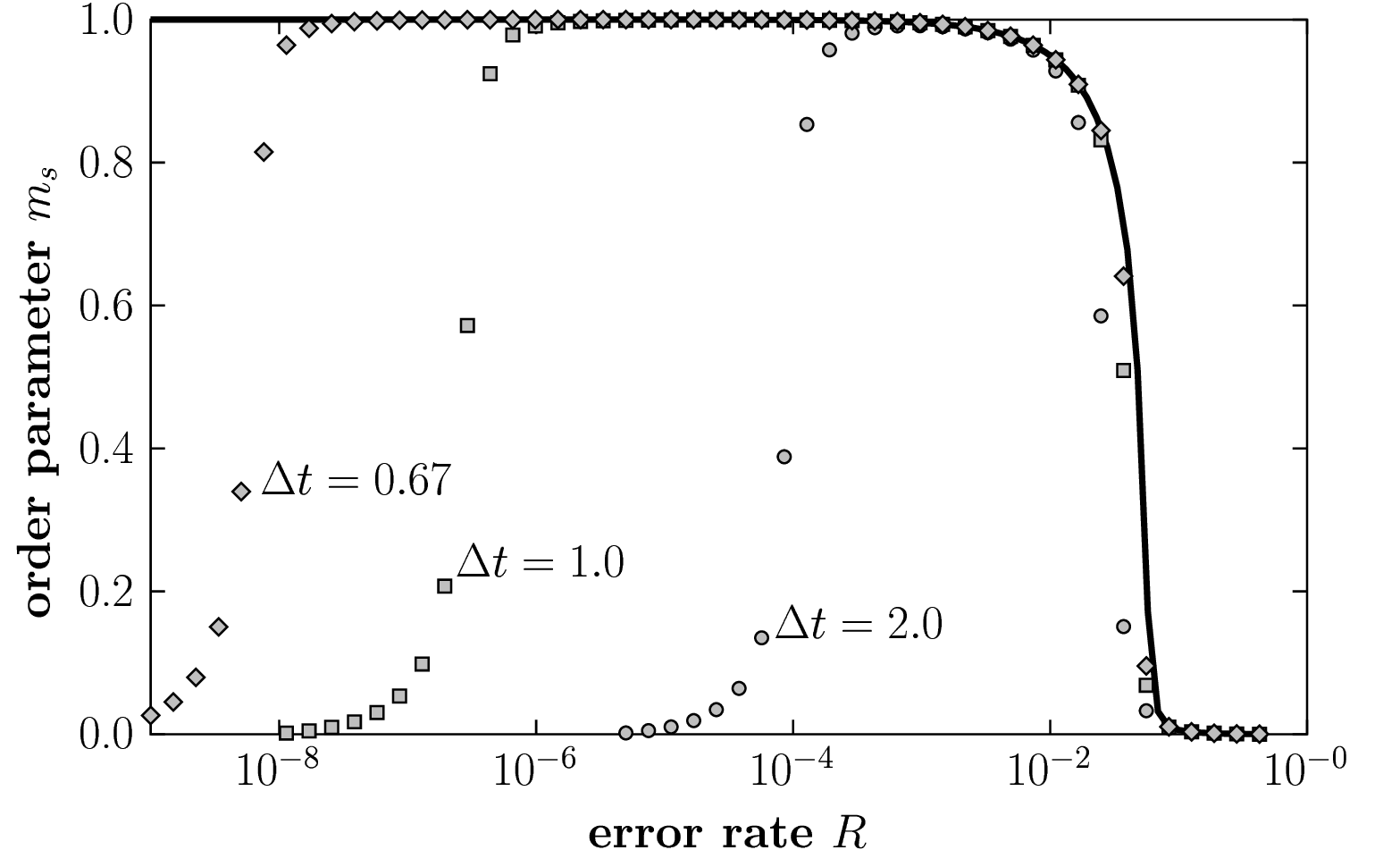}, width=11cm}
}
\caption{Order parameter in a fitness landscape similar to that of
 Fig.~\ref{fig:app-vs-ex}, but with $T=100$. The solid line stems from the
 full continuous time propagator, and the dots have been calculated from the
 discrete approximation Eq.~\eqref{eq:discrete-approx-overl}. The number of
 discretization time steps $n$ is defined as $T/\Delta t$. The graph shows a
 snapshot of the order parameter at phase $\phi=0$ of its limit
 cycle.}\label{fig:alt-master-op}
\end{figure}

As an example, consider Fig.~\ref{fig:alt-master-op}. There we display the
order parameter in the oscillating sharp peak landscape obtained from the full
continuous propagator, and compare it to the result from the discrete
approximation for various values of $\Delta t$. For a relatively large $\Delta
t=2$ ($n=50$), Eq.~\eqref{eq:discrete-approx-overl} gives a poor approximation
of the continuous system. Throughout the whole range of $\mut$ there are
significant deviations from the full solution. As we decrease $\Delta t$
(increase $n$), the approximation moves much closer to the true value of the
order parameter. Yet, for very small $\mut$, the order parameter always decays
to zero in the approximation, whereas it stays close to one in the full
solution. However small we choose $\Delta t$, there will always be some
contribution from the geometric mean at $\mut=0$. That causes the order
parameter in the discrete approximation to vanish for this particular
landscape.

Contrasting to above situation, however, the differences between approximation
and full solution are hardly noticeable in landscapes where the arithmetic and
the geometric mean have a comparable structure (a peak in the averaged
landscape is also a peak in the geometric mean of the geometric mean of the
landscape, only with a slightly different height).

\section{Conclusions}\label{sec:conclusions}

We have studied time-dependent fitness landscapes in the quasispecies model
for the particular regime of small mutation rates. We have shown that the
discrete time formulation and the continuous time formulation yield
qualitatively different outcomes in that regime.  If time is updated
continuously, an evolving population adapts for $\mut\rightarrow 0$ to the
exponential of the average fitness landscape, whereas in discrete time, the
population adapts to the geometric mean of the landscape.

If the arithmetic or the geometric mean of the fitness landscape have
degeneracies, then the behavior of the respective continuous time or discrete
time system for $\mut\rightarrow 0$ is determined by the effect of the
landscape on the population for some small but finite $\mut$, which can be
very different from its effect for $\mut = 0$. In particular, for the case of
a slowly oscillating peak, the growth of the population onto the peak when the
peak is high is much faster than the diffusion away from the peak when the
peak is low, which implies that a population can draw a selective advantage
from that peak \emph{even} if the average (resp.\ geometric mean) height of
the peak does not exceed the surrounding landscape. From that observation,
the following picture emerges: If the average height of a slowly oscillating
peak is larger than or equal to the surrounding fitness landscape, than in a
small mutation rate environment a population will draw a selective advantage
from being close to the peak position. Only if the average height is truly
smaller than the surrounding fitness, the peak position will be necessarily
disadvantageous.

The differences that we have found between continuous time and discrete time
systems are not only interesting from a modeling perspective. They also have
implications for the evolution of organisms that have the ability to influence
their replication cycle. In a fluctuating environment, a strain that feels the
average of the landscape will have a selective advantage over a strain that
feels the geometric mean, as the latter is generally smaller. Hence, if two
strains are identical apart from the fact that one replicates in a
synchronized manner (all individuals generate their offspring at the same
time, every $\Delta t$ units of time), whereas the other one replicates
unsynchronized (at any point in time, some individuals may generate
offspring), then the unsynchronized strain will out-compete the synchronized
strain.

Throughout this paper, we have exclusively considered infinite populations. It
is quite likely that finite populations experience
the arithmetic or geometric mean fitness just as infinite populations do.
However, since finite population sampling occurs at every time step, the
sampling might well interfere with the averaging, such that finite
populations could experience a somewhat different landscape. Nevertheless, the
effect that a fluctuating peak can lead to a selective advantage will also
exist in a finite population. With a small mutation rate, the finite
population will not drift away from the peak when it is below average, and
hence the population will most likely rediscover the peak when it rises again
to above average.

We thank Erik van Nimwegen for useful comments and suggestions, and Chris
Adami for carefully reading the manuscript. This work was supported in part by
the National Science Foundation under contract No.\ DEB-9981397, and by the
BMBF under F\"orderkennzeichen 01IB802C4.


\newpage

\end{document}